\documentclass[twocolumn,twoside]{IEEEtran}

\usepackage{mathpazo}
\usepackage{times}
\usepackage{amsmath}
\usepackage{amsfonts}
\usepackage{latexsym}
\usepackage{amssymb}
\usepackage{cite}

\usepackage{upref}
\usepackage{theorem}
\usepackage{psfrag}
\usepackage{color}

\usepackage[inline]{trackchanges}
\usepackage[pdftex]{graphicx}

\addeditor{}
\addeditor{alex}

\newtheorem{alg}{Algorithm}

\newtheorem{theorem}{Theorem}
\newtheorem{lem}{Lemma}
\newtheorem{clm}[lem]{Claim}
\newtheorem{definition}{Definition}
\newtheorem{corollary}[theorem]{Corollary}
\newtheorem{coro}{Corollary}

\newtheorem{conjecture}{Conjecture}
\newtheorem{proposition}{Proposition}

\newcommand{\BA}{\begin{alg}} \newcommand{\EA}{\end{alg}}
\newcommand{\BE}{\begin{enumerate}} \newcommand{\EE}{\end{enumerate}}
\newcommand{\BT}{\begin{theorem}} \newcommand{\ET}{\end{theorem}}
\newcommand{\BL}{\begin{lem}} \newcommand{\EL}{\end{lem}}
\newcommand{\BCM}{\begin{clm}} \newcommand{\ECM}{\end{clm}}
\newcommand{\BCR}{\begin{coro}} \newcommand{\ECR}{\end{coro}}
\newcommand{\BP}{\begin{proposition}} \newcommand{\EP}{\end{proposition}}

\newcommand{\BI}{\begin{itemize}} \newcommand{\EI}{\end{itemize}}

\def\FullBox{\hbox{\vrule width 8pt height 8pt depth 0pt}}
\newcommand{\qed}{\;\;\;\FullBox}
\newenvironment{prf}{\noindent{\bf Proof:~~}}{\(\qed\)}
\newcommand{\BPF}{\begin{prf}} \newcommand {\EPF}{\end{prf}}
\newenvironment{proofof}[1]{\noindent{\bf Proof of {#1}.~}}{\endprf}
\newcommand{\BPFOF}{\begin{proofof}} \newcommand {\EPFOF}{\end{proofof}}

\newcommand{\BEQN}{\begin{eqnarray}}\newcommand{\EEQN}{\end{eqnarray}}
\newcommand{\BEQ}{\begin{equation}} \newcommand{\EEQ}{\end{equation}}

\newcommand{\eat}[1]{}
\newcommand{\eps}{\varepsilon}

\begin{document}

\title{Minimizing the alphabet size in codes with restricted error sets}
\author{\large
Mira Gonen,
Michael Langberg,
Alex Sprintson
\thanks{Mira Gonen is with the Department of Computer Science, Ariel University, Ariel 40700, Israel (e-mail: mirag@ariel.ac.il).}
\thanks{Michael Langberg is with the Department
 of Electrical Engineering, University at Buffalo (State University of New-York), Buffalo, NY 14260, USA (e-mail: mikel@buffalo.edu). Work supported in part by NSF grant 1909451.}
 \thanks{Alex Sprintson is with the Department of Electrical and Computer Engineering, Texas A\&M University, College Station, TX 77843-3128, USA (e-mail:  spalex@tamu.edu).} 
  %
}



\maketitle

\addtolength{\textfloatsep}{-1.01\baselineskip}
\addtolength{\dbltextfloatsep}{-1.01\baselineskip}
\addtolength{\belowdisplayskip}{-1.01ex}
 \addtolength{\abovedisplayshortskip}{-1.01ex}
 \addtolength{\belowdisplayshortskip}{-1.01ex}
 \renewcommand{\baselinestretch}{0.967}

\begin{abstract}
This paper focuses on error-correcting codes that can handle a predefined set of specific error patterns. The need for such codes arises in many settings of practical interest, including wireless communication and flash memory systems. In many such settings, a smaller field size is achievable than that offered by MDS and other standard codes.   We  establish a connection between the minimum alphabet size for this generalized setting  and the combinatorial properties of a hypergraph that represents the prespecified collection of error patterns. We also show a connection between error and erasure correcting codes in this specialized setting. This allows us to establish bounds on the minimum alphabet size  and show an advantage of non-linear codes over linear codes in a generalized setting. We also consider a variation of the problem which allows a small probability of decoding error and relate it to an approximate version of hypergraph coloring.
\end{abstract}


\section{Introduction}

In many practical settings, there is a need to design error-correcting codes that can handle specific error patterns. For example, in wireless communications, magnetic recording, flash memory systems, and Dynamic Random-Access Memories (DRAMs) the errors can appear in correlated locations such as bursts, single-row errors, or crisscrosss errors, e.g.,~\mbox{\cite{6145514,yaakobi2010error,1090719,MP05,YY19,10.5555/2388996.2389100}}.
These settings benefit from customized error correcting codes, that may improve on the best known parameters of standard error correcting codes.
For example, the optimal error-correcting capabilities of the classical $(n,k)$ Maximum Distance Separable (MDS) code, such as the Reed-Solomon code, come at the price of a significant alphabet size of $q\geq n-k+1$, \cite{B52}.\footnote{The minimum alphabet size of an $(n,k)$ MDS code is unknown, see Conjectures~\ref{con:MDS_new} and \ref{con:MDS_linear_new} in the paper.} As we show in this paper, in many settings with specific error patters,  a much smaller alphabet size is needed.





In this work, we present a general framework for code design that can handle any possible collection of predefined error patterns. Our framework applies to both linear and non-linear codes. 
For an error-correcting code of length $n$, we use an $n$-vertex hypergraph $G$ to represent the given collection of error sets. Specifically, nodes of $G$ represent the coordinates (symbols) of the codewords, while the hyperedges of $G$ represent possible locations for errors, i.e., each hyperedge $e$ represents the set of coordinates that can be corrupted in the specific scenario represented by $e$. For each collection of error sets represented by $G$,  we are interested in finding the minimum alphabet size over which there exists a code that can correct all error sets specified by edges in $G$. 
In our setting, $(n,k)$-MDS codes can correct error patterns corresponding  to the complete $(n-k)/2$-uniform $n$-vertex hypergraph.

In this work, we 
relate the minimum alphabet size of error-correcting codes with predefined error patterns to 
certain variants of hypergraph coloring.
Through reductive arguments to erasure codes, and in particular to our prior work \cite{GHLS20} in the context of erasure codes with generalized decoding sets, we propose code design for the  error setting at hand, and show that non-linear error-correcting codes outperform linear ones.
We then turn
to study a variation of the problem
which allows a small probability of decoding error and relate it to an approximate version of hypergraph coloring.


Our work is structured as follows. In Section~\ref{sec:pre}, we give some preliminaries and, in particular, we introduce our model for generalized erasure and error patterns.  We also 
review our previous study on erasure codes in the generalized setting 
of a predefined collection of decoding sets \cite{GHLS20}. 
In Section~\ref{sec:err},  we present  bounds on the minimum alphabet size of the corresponding codes through hypergraph coloring. In Section~\ref{sec:connection}, we reduce the error-correcting setting to the erasure setting.
In Section~\ref{sec:det}, we extend our studies to the  problem of error detection. 
Finally, in Section~\ref{sec:average}, we relax the zero-error requirement for decoding a correct message and analyze settings which allow small $\eps>0$ probability of decoding error. 

 \section{Model and Preliminaries}
 \label{sec:pre}
 
 
 Since our paper makes a connection between erasure and error correction in a generalized setting, we present definitions for both scenarios. We begin by presenting a definition and our prior results for erasure  correction scenarios.
 
 \subsection{Erasure Correction with predefined decoding sets.}
 
We start by studying the design of erasure-codes in a generalized setting in which decoding is required from a collection of predefined decoding sets. In this setting, the decoding sets include the set of coordinates that can be used to decode the message. The setting is represented by a hypergraph $G=([n],E)$, with the set  $[n]=\{1,\dots,n\}$ of nodes representing coordinates and set of hyperedges $E$ representing decoding sets.

We define the $q_k$ parameter of a given hypergraph $G=([n],E)$ as the minimum alphabet size of a $(n,k)$ erasure code that enables the receiver to decode the original message from every subset $e \in E$. 

\begin{definition}[The $q_k$ parameter \cite{GHLS20}]\label{def:q_parameter}
Let $G=([n],E)$ be a hypergraph on the vertex set $[n]=\{1,\ldots,n\}$.
Let $k$ be integer.
Let $q_k(G)$ denote the smallest size $q$ of an alphabet $F$ for which there exist an encoding function
\[C: F^k \rightarrow F^n\]
and a decoding function
\[ D: (F \cup \{\perp\})^n \rightarrow F^k \]
such that for every edge $e \in E$ and every message $m \in F^k$ it holds that
\[ D(C_e(m)) = m.\]
Here, $C_e(m)$ stands for the word obtained from the codeword $C(m)$ by replacing the symbols in the locations of $[n] \setminus e$ by the erasure symbol $\perp$.

Similarly, let $q_{k,lin}(G)$ denote the smallest prime power $q$ for which there exist {\em linear} encoding and decoding functions defined above when $F$ is a field of size $q$.
\end{definition}

In Definition~\ref{def:q_parameter}, notice that for $G$ that includes edges of size less than $k$ no such $(C,D)$ pair exists (no matter what the size of $F$ is). In this case we define $q_k(G)$ and $q_{k,lin}(G)$ to be $\infty$. Moreover, for every $G$ with edges of size at least $k$, MDS codes satisfy the requirements on $(C,D)$ and thus $q_k(G) < \infty$. 
Specifically, observe that for the complete $n$-vertex $k$-uniform hypergraph, denoted by $\kappa_{n,k}$, the values of $q_k(\kappa_{n,k})$ and $q_{k,lin}(\kappa_{n,k})$  are equal to the minimum alphabet sizes of general and linear $(n,k)$ MDS codes, respectively. We state below the MDS conjectures for general and for linear codes (see, e.g.,~\cite{B52,S55,MDS-linear,Huntemann_thesis}).

\begin{conjecture}[MDS Conjecture for general codes]
\label{con:MDS_new}
For given integers $k < q \neq 6$, let $n(q,k)$ be the largest integer $n$ such that $q_k(\kappa_{n,k}) \leq q$. Then,
\begin{equation}\label{eq:MDS}
n(q,k) \le\left\{
	\begin{array}{ll}
		q+2  & \mbox{if } 4|q \mbox{ and } k\in \{3, q-1\}\\
		q+1 & \mbox{otherwise.}
	\end{array}
\right.
\end{equation}
\end{conjecture}

\begin{conjecture}[MDS Conjecture for linear codes]
\label{con:MDS_linear_new}
For given integers $k < q$ where $q$ is a prime power, let $n(q,k)$ be the largest integer $n$ such that $q_{k,lin}(\kappa_{n,k}) \leq q$. Then,
\begin{equation}\label{eq:MDS}
n(q,k) \le\left\{
	\begin{array}{ll}
		q+2  & \mbox{if } q \mbox{ is even and } k\in \{3, q-1\}\\
		q+1 & \mbox{otherwise.}
	\end{array}
\right.
\end{equation}
\end{conjecture}

%

\noindent

There are strong relations between the $q$ parameter of hypergraphs and certain colorings.
%

\begin{definition}[Hypergraph strong-coloring]\label{def:k_col}
A {\em valid} strong-coloring of a hypergraph $G$ is an assignment of colors to its vertices so that the vertices of each edge are assigned to distinct colors. 
The chromatic number $\chi(G)$ of $G$ is the minimum number of colors that allows a valid strong-coloring of $G$.
At times, we refer to $\chi$ simply as the chromatic number of $G$.
\end{definition}

\begin{definition}[Hypergraph $k$-coloring]\label{def:k_col1}
A {\em valid} $k$-coloring of a hypergraph $G$ is an assignment of colors to its vertices so that the vertices of each edge are assigned to at least $k$ distinct colors.
The $k$-chromatic number $\chi_k(G)$ of $G$ is the minimum number of colors that allows a valid $k$-coloring of $G$. 
If $G$ has edges of size less than $k$, we define $\chi_k(G)=\infty$.
\end{definition}

Note that 
a $k$-coloring of a $k$-uniform hypergraph is exactly a strong-coloring. Also, note that every hypergraph $G$ for which $q_k(G) < \infty$ (i.e., all edges are of size at least $k$) satisfies $\chi_k(G) \leq \chi(G)$.
In particular, for $k$-uniform hypergraphs $G$, $\chi_k(G)=\chi(G)$.

\BT[Connecting $q_k(G)$ with $\chi_k(G)$, \cite{GHLS20}]\label{the:reduction_new}
For every hypergraph $G$ for which $q_k(G) < \infty$,
\[q_k(G) \leq q_k(\kappa_{\chi_k(G),k}) \mbox{ ~~and~~  } q_{k,lin}(G) \leq q_{k,lin}(\kappa_{\chi_k(G),k}).\]
In particular,  \[q_k(G) \leq q_{k,lin}(G) \leq [\chi_k(G)-1]_{pp}.\]
Here, for an integer $x$, $[x]_{pp}$ represents the smallest prime power that is greater or equal to $x$.
\ET

Theorem~\ref{the:reduction_new} formalizes the natural intuition that for {\em simple} collections of erasure patterns $G$, i.e., the setting in which  $\chi_k(G)$ is small, a {\em small} alphabet size $q$ suffices for a suitable erasure code. In particular, the theorem states that $q_k(G)$ is upper bounded by  $q_k(\kappa_{\chi_k(G),k})$, which is the minimum alphabet size of a $(\chi_k(G),k)$ MDS code.

The graph family $G_{q,k}$, defined next, is helpful in analyzing the tightness of the  upper bound provided by Theorem~\ref{the:reduction_new}.

\begin{definition}[The graph family $G_{q,k}$]
\label{def:Hqk}
For integers $q$ and $k$, let $G_{q,k}$ be the $k$-uniform hypergraph whose vertex set consists of all the balanced vectors of length $q^k$ over $F=\{0,1,\dots,q-1\}$, that is, the vectors $u \in F^{q^k}$ such that $|\{i \in [q^k] \mid u_i =j\}|=q^{k-1}$ for every $j \in F$, where $k$ vertices $u^1=(u^1_1,\dots,u^1_{q^k}),...,u^k=(u^k_1,\dots,u^k_{q^k})$ form an edge if the collection of $k$-tuples $\{(u^1_i,u^2_i,\ldots,u^k_i)\}_{i \in [q^k]}$ is equal to $[q]^k$.
\end{definition}


The following lemma identified hypergraphs $G$ for which the gap between $q_k(G)$ and $\chi_k(G)$ is maximal.

\BL[The extremal nature of $G_{q,k}$, \cite{GHLS20}]
\label{lem:subgraph_new}
For integers $q$ and $k$,
\begin{enumerate}
  \item $q_k(G_{q,k}) \leq q$, and
  \item $\chi_k(G) \leq \chi_k(G_{q,k})$ for every graph $G$ with $q_k(G)=q$.
\end{enumerate}
\EL

Extending results in \cite{GHLS20}, below we present (rater loose) bounds on $\chi(G_{q,k})$.
\begin{proposition} [Bounds of $\chi_k(G_{k,q})$]
 \label{prop:chi}
 For every prime power $q$ and $k\ge 2$,
 $$
 \frac{q^k-1}{q-1}\le\chi_k(G_{q,k})\le {q^{k-1}+1\choose q^{k-2}+1}.
 $$
 \end{proposition}
  \begin{prf}
  We first study the collection of vertices in $G_{q,k}$ corresponding to normalized linear functions $F^k \rightarrow F$ for field $F$ of size $q$.  A normalized linear function is one in which the leading nonzero coefficient equals 1. Such functions, when considered in vector form  $(u_1,\ldots u_{q^k})\in F^{q^k}$ are balanced and thus correspond to vertices of $G_{q,k}$. Moreover,  it is not hard to verify that any two vertices of $G_{q,k}$ corresponding to distinct normalized linear functions are included in an edge of $G_{q,k}$ (i.e., there exist $k-2$ additional vertices of $G_{q,k}$ corresponding to normalized linear functions that complete a linearly independent collection of functions). Thus, any $k$-coloring of $G_{q,k}$ must color all vertices corresponding to normalized linear functions with distinct colors. The number of  normalized linear functions over $F$ corresponding to vertices of $G_{q,k}$ is $\sum_{i=1}^k{q^{k-i}}=(q^k-1)/(q-1)$. Therefore $\chi(G_{q,k})\ge \frac{q^k-1}{q-1}$. 

 On the other hand, we now show that $\chi(G_{q,k}) \leq {q^{k-1}+1\choose q^{k-2}+1}$.
  For any vector $u$ in $G_{q,k}$, consider the first $q^{k-1}+1$ entries of $u$. By the pigeonhole principal, $u_{i_1} = u_{i_2}=\ldots =u_{i_{q^{k-2}+1}}$ for some collection of entries indexed by $i_1<i_2<\ldots<i_{q^{k-2}+1} \in [q^{k-1}+1]$. 
  Now, for any $q^{k-2}+1$ distinct indices $i_1<i_2<\ldots<i_{q^{k-2}+1} \in [q^{k-1}+1]$ let $A_{i_1,i_2,\ldots,i_{q^{k-2}+1}}$ be the set of all vertices $u \in F^{q^k}$ of $G_{q,k}$ that satisfy $u_{i_1} = u_{i_2}=\ldots =u_{i_{q^{k-2}+1}}$. Every set $A_{i_1,i_2,\ldots,i_{q^{k-2}+1}}$ forms an independent set in $G_{q,k}$, i.e., a set that does not include any two vertices from an edge of $G_{q,k}$.
 This follows, since for every two distinct vertices $u,v \in A_{i_1,i_2,\ldots,i_{q^{k-2}+1}}$ we have $u_{i_1} = u_{i_2}=\ldots =u_{i_{q^{k-2}+1}}$ and $v_{i_1} = v_{i_2}=\ldots =v_{i_{q^{k-2}+1}}$, 
  which is too large of an overlap to allow the balanced nature of vertices included in edges of $G_{q,k}$.
  Specifically, for vertices $u$ and $v$ that appear in an edge of $G_{q,k}$, it must be for any $j_1$ and $j_2$ in $F$ that $|\{i \in [q^k]| u_i=j_1, v_i=j_2]\}|=q^{k-2}$.
  Such independent sets were referred to as {\em canonical} in \cite{GHLS20}.
  As the $q^{k-1}+1 \choose q^{k-2}+1$  independent sets $A_{i_1,i_2,\ldots,i_{q^{k-2}+1}}$ of $G_{q,k}$ with $i_1,i_2,\ldots,i_{q^{k-2}+1} \in [q^{k-1}+1]$ cover the entire vertex set of $G_{q,k}$, coloring each one with a distinct color implies the required upper bound on $\chi_k$.
 \end{prf}

Lemma~\ref{lem:subgraph_new} and Proposition~\ref{prop:chi} 
imply a gap between $q_k(G_{q,k})$ and $\chi_k(G_{q,k})$ which can be extended to one between $q_{k,lin}$ and 
the $k$-chromatic number of
the subgraph of $G_{q,k}$ induced by vertices that correspond to normalized linear functions. 
\BP[Gap between $q_{k,lin}(G)$ and $\chi_k(G)$, \cite{GHLS20}]
\label{prop:GL_new}
For every $k\ge 3$ and every prime power $q$, there exists a $k$-uniform hypergraph $G$ with $q_{k,lin}(G) \leq q$ and yet $\chi_k(G) \geq \frac{q^k-1}{q-1}$.
\EP

%

We finally state a modest known gap between $q_{k,lin}$ and $q_k$. Identifying graphs that exhibit a larger gap than that presented below is a problem left open in this work. 
\BP[Gap between $q_{k,lin}$ and $q_k$, \cite{LL05}]
\label{prop:gap}
For $q=3$ and $k=2$ it holds that 
$$
q_{k,lin}(G_{q,k}) = [\chi_k(G_{q,k})-1]_{pp} =5 > 3 \geq q_k(G_{q,k}).
$$
\EP

 \subsection{Error Correction with predefined error sets.}

In what follows, we extend our discussion beyond erasures to the context of {\em errors}.
As we will see, several of our results on the $q$-parameter corresponding to erasures extend naturally to the $p$-parameter (defined below) corresponding to codes with restricted error sets. Similarly, to the erasure setting, we represent the collection of error sets by using a hyper-graph $G=([n],E)$, in which the set of vertices $[n]$ represents coordinates of a codeword. Each edge $e\in E$ of $G$ represents an error set, i.e., the set of the coordinates that can be altered.  Note that this is different from the notation used in  Definition~\ref{def:q_parameter} for the erasure case in which edges $e$ represented decoding sets (i.e., sets of uncorrupted symbols).


\begin{definition}[The $p_k$ parameter]\label{def:p_parameter}
Let $G=([n],E)$ be a hypergraph on the vertex set $[n]=\{1,\ldots,n\}$.
Let $k$ be an integer.
Let $p_k(G)$ denote the smallest size $p$ of an alphabet $F$ for which there exist an encoding function
\[C: F^k \rightarrow F^n\]
and a decoding function
\[ D: F^n \rightarrow F^k \]
such that for every edge $e \in E$, every message $m \in F^k$, and every error vector $v=(v_1,\dots,v_n) \in F^n$,
\[ D(C(m) \diamond_e v) = m.\]

Here, for $C(m)=c_1,\dots,c_n$, the term $C(m) \diamond_e v$ refers to the vector $y=y_1,\dots, y_n$ for which for $i \in [n]$, $y_i=v_i$ if $i\in e$, and otherwise $y_i=c_i$ (i.e., we overwrite $C(m)$ with values of $v$ in the coordinates $i \in e$).

Similarly, let $p_{k,lin}(G)$ denote the smallest prime power $p$ for which there exist {\em linear} encoding and decoding functions as above when $F$ is a field of size $p$.
\end{definition}

In Definition~\ref{def:p_parameter}, the pair $(C,D)$ corresponds to a code that is resilient to errors on locations corresponding to an edge $e\in E$.
That is, the edge set $E$ represents the possible error patterns (i.e., sets of potentially corrupted symbols).

Similar to Definition~\ref{def:q_parameter}, in Definition~\ref{def:p_parameter}, if $G$ has edges of size greater than $\left\lfloor\frac{n-k}{2}\right\rfloor$, no such codes $(C,D)$ exist, and we define $p_k(G)=p_{k,lin}(G)=\infty$.

As with erasures, for the complete hypergraph $\kappa_{n,\left\lfloor\frac{n-k}{2}\right\rfloor}$, the values of $p_k(\kappa_{n,\left\lfloor\frac{n-k}{2}\right\rfloor})$ and $p_{k,lin}(\kappa_{n,\left\lfloor\frac{n-k}{2}\right\rfloor})$ are equal to the minimum alphabet sizes of general and linear $(n,k)$ MDS codes, respectively. That is, $p_k(\kappa_{n,\left\lfloor\frac{n-k}{2}\right\rfloor})=q_k(\kappa_{n,k})$ and $p_{k,lin}(\kappa_{n,\left\lfloor\frac{n-k}{2}\right\rfloor})=q_{k,lin}(\kappa_{n,k})$.

Note that  Definitions~\ref{def:q_parameter} and \ref{def:p_parameter} assume zero-error decoding. We relax this requirement in Section~\ref{sec:average}.

\section{Bounds on the Alphabet Size}
\label{sec:err}

\begin{proposition}[Analog of Theorem~\ref{the:reduction_new}]\label{prop:p-upper bound col}
Let $k$ be an integer.
For every hypergraph $G=([n],E)$ for which $p_k(G)<\infty$ 
it holds that $$p_k(G)\le p_k(\kappa_{\chi,\left\lfloor\frac{\chi-k}{2}\right\rfloor}),$$ where $\chi=\chi(\bar{G})$ and  $\bar{G}=(V,\bar{E})$ is the hypergraph with vertex set $V= [n]$ and edges $\bar{E}=\{V \setminus e | e \in E\}$.
\end{proposition}

\begin{prf}
To ease our notation, we assume that $n-k$ and $\chi - k$ are even (minor modifications in notation are needed otherwise).
Let $G$ be as above and let $\chi=\chi(\bar{G})$.
Denoting $p = p_k(\kappa_{\chi,(\chi-k)/2})$, it follows  that there exist a $(\chi,k)$ MDS code $C$ over an alphabet $F$ of size $p$.
To prove that \mbox{$p_k(G) \leq p$}, we define a coding scheme for $G$ over the alphabet $F$ that includes the following two steps.
First, fix a valid coloring $g:[n] \rightarrow [\chi]$ of $\bar{G}$.
Second, consider the encoding function $\widetilde{C}:F^k \rightarrow F^n$ that given a message $m \in F^k$ outputs the vector in $F^n$ whose $i$'th entry $\widetilde{C}_i(m)$ is $C_{g(i)}(m)$, i.e., $\widetilde{C}_i(m)$ is the coordinate in the codeword $C(m)$ which corresponds to the color of the $i$'th vertex. Here, and throughout, we use the notation $C_i(m)$ to denote the $i$'th entry in the codeword $C(m)$.

The decoder $\widetilde{D}:F^n \rightarrow F^k$ for $G$ is now defined using the following procedure.
Consider an error vector $v\in F^n$, edge $e_0 \in E$, 
and the corresponding received word $y=\widetilde{C}(m) \diamond_{e_0} v$.
For each edge $\bar{e}$ in $\bar{E}$, the decoder $\widetilde{D}$ considers $y_{\bar{e}}$ consisting of the entries of $y$ restricted to the indices in ${\bar{e}}$, and detects whether $y_{\bar{e}}$ has been corrupted, i.e., whether $\widetilde{C}_{\bar{e}}(m)=y_{\bar{e}}$.
As for at least one such edge $\bar{e}_0$ it holds that $\widetilde{C}_{\bar{e}_0}(m)=y_{\bar{e}_0}$  (e.g. for $\bar{e}_0 = [n]\setminus {e}_0$), the decoder $\widetilde{D}$ can use $y_{\bar{e}_0}$ to decode $m$.
We are left to show, given $\bar{e} \in \bar{E}$,  how $\widetilde{D}$ can detect whether $\widetilde{C}_{\bar{e}}(m)=y_{\bar{e}}$, and if so decode $m$.

To detect whether a given $\bar{e}$ in $\bar{E}$ satisfies $\widetilde{C}_{\bar{e}}(m)=y_{\bar{e}}$ we note, by the definition of $\widetilde{C}$ and the fact that all vertices in $\bar{e}$ have distinct colors under the coloring $g$, that the entries in $\widetilde{C}_{\bar{e}}(m)$ correspond to at least $(n+k)/2$ distinct entries $C(m)$. The latter, in turn, implies that  $\widetilde{C}_{\bar{e}}(m)$ is itself a $(|\bar{e}|,k)$ MDS code. As such, $\widetilde{C}_{\bar{e}}(m)$ can detect up to $|\bar{e}| - k \geq \frac{n-k}{2}$ errors and correct up to $\left(|\bar{e}| - k\right)/2 \geq \frac{n-k}{4}$ errors.  We conclude, as all error sets $e$ are of size at most $(n-k)/2$, that given $\bar{e}$ in $\bar{E}$, the decoder $\widetilde{D}$ can detect whether or not $y_{\bar{e}}$ has been corrupted, and if not, recover $m$ as required.
\end{prf}

 Proposition~\ref{prop:p-upper bound col} is not tight,  meaning that $p_k(G)$ might be smaller than $p_k(\kappa_{\chi,(\chi-k)/2})$. For $k=2$ take for example  $G=([6],E)$ to be the $6$-cycle, i.e., the graph on $6$ vertices in which its edges $E=\{(i,i+1)| i =0,1,\dots,5\}$ (with addition $\mod$ 6). Then $p_2(G)=2$, since the binary encoding $C: F^2 \rightarrow F^6$ in which for a message $m=(x,y)\in F^2$ equals  $C(x,y)=(x,y,x,y,x,y)$ allows majority decoding for any 2 errors along an edge in $G$. However, $\chi=\chi(\bar{G})=6$, since every pair of vertices in $\bar{G}$ is included in some edge in $\bar{E}$, and by~\cite{H12} it holds that $p_2(\kappa_{\chi,\left\lfloor\frac{\chi-k}{2}\right\rfloor})=p_2(\kappa_{6,2})=5$.
 In the next section, we improve on Proposition~\ref{prop:p-upper bound col} by connecting the $p_k$ and $q_k$ parameters.

\section{Connecting Error and Erasure Correcting Codes}
\label{sec:connection}

For parameters $n$ and $k$, we say that encoder $C: F^k \rightarrow F^n$ is  {\em good} for a given hypergraph $G=([n],E)$ with respect to erasures (res., errors) if there exists a decoder $D$ satisfying Definition~\ref{def:q_parameter} (res., Definition~\ref{def:p_parameter}).
The following proposition is proven from basic principles.

\begin{proposition}[From errors to erasures]
\label{prop:errors}
Let $n$ and $k$ be parameters. Consider a hypergraph $G^{\tt err}=([n],E^{\tt err})$ corresponding to errors. Let $G^{\tt era}=([n],E^{\tt era})$ be the hypergraph (corresponding to erasures) for which
$$
E^{\tt era}= \{[n] \setminus (e_1^{\tt err} \cup e_2^{\tt err}) \mid e_1^{\tt err},e_2^{\tt err} \in E^{\tt err}\}.
$$
Let $C: F^k \rightarrow F^n$ be any encoder.
Then, $C$ is good for $G^{\tt err}$ if and only if $C$ is good for $G^{\tt era}$.
 \end{proposition}
 \begin{prf}
First assume that  $C$ is good for $G^{\tt err}$. We show that for every edge $e=e^{\tt era} \in E^{\tt era}$, one can decode $m$ from $C_e(m)$. Assume in contradiction that there are two messages $m_1 \neq m_2$ such that $C_e(m_1)=C_e(m_2)$.
Recall that $e=[n] \setminus (e_1 \cup e_2)$ for $e_1= e_1^{\tt err} \in  E^{\tt err}$ and $e_2 = e_2^{\tt err} \in  E^{\tt err}$.
Consider the word $y=(y_1,\dots,y_n) \in F^n$ such that for $i \in e= [n] \setminus (e_1 \cup e_2)$: $y_i = C_i(m_1)=C_i(m_2)$, for $i \in e_1 \setminus e_2$: $y_i = C_i(m_2)$, and for $i \in e_2$: $y_i = C_i(m_1)$.
It is not hard to verify that there exist vectors $v_1$ and $v_2$ such that \mbox{$y=C(m_1) \diamond_{e_1} v_1=C(m_2) \diamond_{e_2} v_2$}.
Namely, $y$ could be obtained from the codeword $C(m_1)$ with error vector $v_1$ corresponding to $e_1$ or from the codeword $C(m_2)$ with error vector $v_2$ corresponding to $e_2$, contradicting the existence of a decoder $D$ according to Definition~\ref{def:p_parameter}.


For the other direction, if code $C$ is not good for $G^{\tt err}$ then there exist two messages, $m_1$ and $m_2$,  two error vectors $v_1$ and $v_2$,  and two edges $e_1$ and $e_2$ in $E^{\tt err}$ such that  
$
C(m_1) \diamond_{e_1} v_1=C(m_2) \diamond_{e_2} v_2.
$
Otherwise, it is not hard to verify the existence of a natural decoder $D$ according to Definition~\ref{def:p_parameter}.
Let $e=[n] \setminus (e_1 \cup e_2) \in E^{\tt era}$. The equality $C(m_1) \diamond_{e_1} v_1=C(m_2) \diamond_{e_2} v_2$ now implies that
$C_e(m_1)=C_e(m_2)$, which in turn implies that $C$ is not good for $G^{\tt era}$.
 \end{prf}

The proposition above has an operational perspective.
Namely, one can design an error-correcting code $C$ and decoder $D$ for a given graph $G^{\tt err}$, by designing an erasure-code for  the graph $G^{\tt era}$. The latter can be done, e.g., using   Theorem~\ref{the:reduction_new} to obtain the following corollary.
\begin{corollary}\label{cor:p-upper bound col}
Let $k$ be an integer.
For every hypergraph $G^{\tt err}=([n],E)$ for which $p_k(G^{\tt err})<\infty$ 
it holds that 
$$p_k(G^{\tt err})\le q_k(G^{\tt era}) \leq
q_k(\kappa_{\chi_k(G^{\tt era}),k}) \leq
[\chi_k(G^{\tt era})-1]_{pp},$$

which, in turn, implies that 
$$p_k(G^{\tt err})\le 
p_k(\kappa_{\chi,\left\lfloor\frac{\chi-k}{2}\right\rfloor}),$$

where $\chi=\chi_k(G^{\tt era})$.

\end{corollary}

We now extend the connections implied by Proposition~\ref{prop:errors} to capture the $p_k$ and $q_k$ parameters.

\begin{theorem}[Connecting $p_k$ with $q_k$]
\label{thm:errors}
Let $n,k$ be parameters such that $n-k \geq k$.
Let $G_0^{\tt era}=([n],E_0^{\tt era})$ be a hypergraph corresponding to erasures such that $q_k(G_0^{\tt era})<\infty$.
Then, for $N=2n-k$ there exists a hypergraph $G^{\tt err}$ on $N$ vertices such that  $p_k(G^{\tt err})=q_k(G_0^{\tt era})$ and $p_{k,lin}(G^{\tt err})=q_{k,lin}(G_0^{\tt era})$.
\end{theorem}

\begin{prf}
Let $G_0^{\tt era}=([n],E_0^{\tt era})$ be as above.
We define two graphs according to $G_0^{\tt era}$.
First consider the graph $G^{\tt err}=([n] \cup U,E^{\tt err})$ corresponding to errors for which $U$ is a vertex set of size $n-k$ and
$$
E^{\tt err}=\{U\} \cup \{[n] \setminus e^{\tt era} | e^{\tt era} \in E_0^{\tt era}\}.
$$
Here, we use the fact that edges in $E_0^{\tt era}$ are subsets of $[n]$.
Namely, the vertex set $[n] \cup U$ of $G^{\tt err}$ is of size $N=2n-k$ and each edge in $E^{\tt err}$ is of size at most $\frac{N-k}{2}=n-k$.
We refer to the edges in $\{[n] \setminus e^{\tt era} | e^{\tt era} \in E_0^{\tt era}\} \subset E^{\tt err}$ as {\em ordinary} edges, and to the edge $U \in E^{\tt err}$ as the {\em special} edge.

Let $G^{\tt era}$ be the graph corresponding to erasures defined by $G^{\tt err}$ as in Proposition~\ref{prop:errors}.
Namely, $G^{\tt era}=([n] \cup U,E^{\tt era})$ where
$$
E^{\tt era}= \{([n] \cup U) \setminus (e_1^{\tt err} \cup e_2^{\tt err}) \mid e_1^{\tt err},e_2^{\tt err} \in E^{\tt err}\}.
$$
Taking a closer look at the edge set $E^{\tt era}$, if an edge $e^{\tt era}$ in $E^{\tt era}$ is defined by two ordinary edges of $E^{\tt err}$, then  it is not hard to verify that $U \subseteq e^{\tt era}$. If an edge $e^{\tt era}$ in $E^{\tt era}$ is defined by the special edge $U$ and an ordinary edge $e \in E^{\tt err}$, then $e^{\tt era} = [n]\setminus e$. As the ordinary edge $e \in E^{\tt err}$, by definition, equals $[n]\setminus e_0^{\tt era}$ for an edge $e_0^{\tt era} \in E_0^{\tt era}$ we conclude that $e^{\tt era} = e_0^{\tt era}$.
Finally, if an edge $e$ in $E^{\tt era}$ is defined solely by $U$ (i.e., we set $e_1=e_2=U$), then $e=[n]$.
All in all, we conclude that the edge set  $E^{\tt era}$ equals the edges $E_0^{\tt era} \cup \{[n]\}$ and an additional set of edges $e^{\tt era}$ for which $U \subseteq e^{\tt era}$.

We now show that $p_k(G^{\tt err})=q_k(G_0^{\tt era})$.
We start by studying codes for $G_0^{\tt era}$ and $G^{\tt era}$.
For any code $C_0: F^k \rightarrow F^n$ for $G_0^{\tt era}$, define the code $C: F^k \rightarrow F^{n+(n-k)}$ for $G^{\tt era}$ in which for every message $m$ it holds that $C(m)=C_0(m)$ on the first $[n]$ entries, that $C(m)=m$ on entries $n+1,\dots,n+k$ and that $C(m)$ equals the symbol $a \in F$ for the remaining entries $n+k+1,\ldots, 2n{-}k$. Here, we use the fact that $n-k \geq k$.
Similarly, for any code $C: F^k \rightarrow F^{n+n-k}$ for $G^{\tt era}$, let the code $C_0: F^k \rightarrow F^{n}$ for $G_0^{\tt era}$ be the restriction of $C$ to the first $n$ entries.
It is now not hard to verify that $C_0$ is good for $G_0^{\tt era}$ if and only if $C$ is good for $G^{\tt era}$.
More specifically, let $C_0$ be a code that is good for $G_0^{\tt era}$, and let $D_0$ be the corresponding decoder. For any message $m$ and edge $e=e_0^{\tt era}$ it holds that $D_0((C_0)_e(m))=m$.
To show that $C$ is good for $G^{\tt era}$ we define the decoder $D$, that for $e^{\tt era} \in E^{\tt era}$ either runs $D_0$ on the first $n$ entries of $C$ if  $e^{\tt era} \subseteq [n]$, or decodes using the identity mapping from $U$ if $U \subseteq e^{\tt era}$.
For the opposite direction, let $C$ be good for $G^{\tt era}$, and let $D$ be the corresponding decoder.
To show that $C_0$ is good for $G_0^{\tt era}$ we define the decoder $D_0$ as the restriction of $D$ that takes into account only the first $n$ entries of $C$. Correctness follows as $E_0^{\tt era} \subseteq E^{\tt era}$ and as $C_0$ is a restriction of $C$ to the first $n$ entries.

To show that $p_k(G^{\tt err})=q_k(G_0^{\tt era})$, let $N=2n-k$ and let $C: F^k \rightarrow F^N$ be any encoder.
By Proposition~\ref{prop:errors}, $C$ is good for $G^{\tt err}$ if and only if  $C$ is good for $G^{\tt era}$. By the discussion above, $C$ is good for $G^{\tt era}$ if and only if the corresponding $C_0$ is good for $G_0^{\tt era}$.
Thus, $C_0$ is good for $G_0^{\tt era}$ if and only if $C$ is good for $G^{\tt err}$.
Optimizing over $|F|$, we conclude $p_k(G^{\tt err})=q_k(G_0^{\tt era})$.
As the reductions described above between $C$ and $C_0$ preserves linearity, we also conclude that $p_{k,lin}(G^{\tt err})=q_{k,lin}(G_0^{\tt era})$.
\end{prf}



By Theorem~\ref{thm:errors} the gap between the $q_k$ parameter and the $q_{k,lin}$ parameter  for erasure codes stated in Proposition~\ref{prop:gap} implies a gap between the $p_k$ parameter and the $p_{k,lin}$ parameter for error correcting codes. We summarize this results in the following corollary. 

\begin{corollary}[Non-linear codes outperform linear codes]
For $k=2$, there exists a hypergraph $G$ with $p_{k,lin}(G) = 5$ and yet $p_k(G)=3$.
\end{corollary}

\section{Error detection}
\label{sec:det}

Similar to the case of errors and erasures, one can define analogs of Definitions~\ref{def:q_parameter} and ~\ref{def:p_parameter} for the case of error detection. Namely, for a given hypergraph $G=([n],E)$ the $r_k$ parameter defined below equals the minimum size alphabet of an $(n,k)$ error detection code that can detect error patters represented by $E$.

\begin{definition}[The $r_k$ parameter]\label{def:r_parameter}
Let $G=([n],E)$ be a hypergraph on the vertex set $[n]=\{1,\ldots,n\}$, and let $k$ be an integer.
Let $r_k(G)$ denote the smallest size $r$ of an alphabet $F$ for which there exist an encoding function
\[C: F^k \rightarrow F^n\]
and a decoding function
\[ D: F^n \rightarrow \{\text{\em error, no-error}\} \]
such that for every edge $e \in E$, every message $m \in F^k$, and every error vector $v=(v_1,\dots,v_n) \in F^n$,
\[D(C(m) \diamond_e v) = ``\text{\em error}'' \ \text{\em{if and only if}}\  C(m)\neq C(m) \diamond_e v,  
\]
($\diamond_e$ is defined in Definition~\ref{def:p_parameter}).



\end{definition}

Similar to Definition~\ref{def:q_parameter}, in Definition~\ref{def:r_parameter}, $r_k(G)$ is defined if and only if all edges of $G$ are of size at most $n-k$, otherwise we define $r_k(G)=\infty$.
Also, similar to Proposition~\ref{prop:errors}, the following proposition is proven from basic principles (its proof is sketched here for completeness).


\begin{proposition}[Detecting errors vs. correcting erasures]
\label{prop:era-detecting}
Let $n$ and $k$ be parameters such that $n-k \geq k$.
For a hypergraph $G=([n],E)$, let $\bar{G}=([n],\bar{E})$ be the hypergraph for which $\bar{E}=\{[n]\setminus e |e\in E\}$.
Then, $r_k(G)=q_k(\bar{G})$.



\end{proposition}

\begin{prf}
Assume that  $\bar{C}$ is a good erasure code for $\bar{G}$. 
The same code can be used for detection on $G$. 
Namely, given a received word $y$, to check if $y$ is corrupted in locations corresponding to $e \in E$, decode to $m$ using $y_{\bar{e}}$ (via the erasure decoding) and compare $\bar{C}(m)$ to $y$.
For the other direction, assume that $C$ is a good detection code for $G$. Use the same code $C$ for erasures. To decode from $C_{\bar{e}}(m)$, construct the collection $Y$ of size $|F|^{n-|{\bar{e}}|}$ of words $y \in F^n$ that equal $C_{\bar{e}}(m)$ on the locations of ${\bar{e}}$ and otherwise equal a (distinct) word in $F^{n-|{\bar{e}}|}$. As $C$ is a detection code for errors with support $e = [n] \setminus \bar{e}$, we can detect the unique $y \in Y$ that is a codeword, and accordingly decode $m$.
 \end{prf}
 

\section{Average error $\varepsilon$}
\label{sec:average}

In what follows, we generalize the $q_k$, $p_k$, and $r_k$ parameters to include a decoding error. 
In our prior work \cite{GHLS20}, for $k=2$ in the context of erasures, we considered decoding error when averaged over the message set $F^k$. 
We here consider a looser notion of error that is also averaged over edges in the edge set $E$ of the hypergraph at hand. 
As shown below, allowing a slight error in decoding will in turn allow the construction of codes with small alphabet sizes (independent of the blocklength $n$). 


\begin{definition}[The $q_{\eps,k}$, $p_{\eps,k}$, and $r_{\eps,k}$ parameters]\label{def:q_avr_eps_parameter}
Let $k$ be an integer.
Let $G=([n],E)$ be a hypergraph on the vertex set $[n]$ and let $\eps >0$.
Let $q_{\eps,k}(G)$ denote the smallest size $q$ of an alphabet $F$ for which there exist an encoding function
$C: F^k \rightarrow F^n$
and a decoding function
$D: (F \cup \{\perp\})^n \rightarrow F^k$
such that 
\[\Pr_{e,m}[D(C_e(m))=m] \geq 1-\eps,\]
where $m$ is uniformly chosen from $F^k$, and $e$ is uniformly chosen from $E$.
One may define $p_{\eps,k}(G)$ and $r_{\eps,k}(G)$ in an analogous manner.
\end{definition}

We will need the following {\em approximate} version of coloring.


\begin{definition}[Hypergraph $(1-\varepsilon)$-$k$-coloring]
\label{def:app_k_col}
A {\em valid} $(1-\varepsilon)$-$k$-coloring of a hypergraph $G=(V,E)$ is an assignment of colors to its vertices $V$ so that for at least $(1-\varepsilon)|E|$ edges $e \in E$, the vertices of $e$ are assigned to at least $k$ colors.
The $(1-\varepsilon)$-$k$-chromatic number $\chi_{\varepsilon,k}(G)$ of $G$ is the minimum number of colors that allows a valid $(1-\varepsilon)$-$k$-coloring of $G$.
\end{definition}

\BT\label{the:reduction_new_avr_eps}
\label{the:average}
Let $G=(V,E)$ be a hypergraph, and $\eps>0$ a parameter.
Then $q_{\varepsilon,k}(G) \leq [\chi_{\varepsilon,k}(G)-1]_{pp}$.
In particular, let $n$ and $k$ be any integers,
$q_{\eps,k}(\kappa_{n,k}) \leq O(k^2/\varepsilon)$.
%
\ET

\begin{prf}
The proof that $q_{\varepsilon,k}(G) \leq [\chi_{\varepsilon,k}(G)-1]_{pp}$ is almost identical to the proof of Theorem~\ref{the:reduction_new} (presented in \cite{GHLS20}) and is obtained by replacing $q_k$ and $\chi_k$ by $q_{\varepsilon,k}$ and $\chi_{\varepsilon,k}$ respectively.
The second part of the theorem follows by showing that $\kappa_{n,k}$ can be $(1-\varepsilon)$-$k$ colored using $k^2/\varepsilon$ colors. Consider partitioning $[n]$ into $k^2/\varepsilon$ subsets, each of size $\eps n/k^2$. Assign the same color to all the vertices in the same subset, and distinct colors to vertices in distinct subsets. We now show that this is a $(1-\varepsilon)$-$k$ coloring. The fraction of  edges that are  assigned to at least $k$ colors is $$\frac{{k^2/\varepsilon \choose{k}}\cdot (\eps n/k^2)^k}{{n\choose{k}}}.$$


Now, for integers $a$ and $b$, ${a\choose b} = \frac{\prod_{j=0}^{b-1}(a-j)}{b!}$, and $a^b\ge \prod_{j=0}^{b-1}(a-j) \ge (a-b)^b\ge (1-b^2/a)a^b$,  thus  
$$
\frac{(1-b^2/a)a^b}{b!}\le {a\choose b}\le \frac{a^b}{b!}.
$$
Therefore $\frac{{k^2/\varepsilon \choose{k}}\cdot (\eps n/k^2)^k}{{n\choose{k}}}\ge 1-\eps$.
\end{prf}

Notice that implications corresponding to those in Theorem~\ref{the:average} on parameters $p_{\eps,k}$ and $r_{\eps,k}$ can be derived using Theorem~\ref{thm:errors} and Proposition~\ref{prop:era-detecting}, respectively.

\bibliographystyle{unsrt}
\bibliography{CodedCooperative31}

\end{document}